\documentclass[9pt,twocolumn]{article}
\usepackage{multirow}
\usepackage{graphicx}
\usepackage{titling}
\usepackage{authblk}
\title{$\theta$-curves in proteins}
\author[1,2,3]{Pawel~Dabrowski-Tumanski}
\author[3,4]{Dimos~Goundaroulis}
\author[3,4]{Andrzej~Stasiak}
\author[1,2]{Joanna~I.~Sulkowska*}
\affil[1]{Faculty of Chemistry, University of Warsaw, Poland}
\affil[2]{Center of New Technologies, University of Warsaw, Poland}
\affil[3]{Center for Integrative Genomics, University of Lausanne, 1015 Lausanne, Switzerland}
\affil[4]{Swiss Institute of Bioinformatics, 1015 Lausanne, Switzerland}
\date{}                     
\begin{document}

\twocolumn[
  \begin{@twocolumnfalse}
  \maketitle
  \begin{abstract}
  Apart from the knots formed by the main-chain, the proteins can form numerous topological structures, when included the covalent and ion-mediated interactions. In this work, we define the protein non-trivial $\theta$-curves and identify 7 different topologies in all structures known up to date. We study the correlation of the motif with the function and organism of origin, and pointing the similarity with main-chain knots, we show that some motifs may indeed be functional. We also analyze the folding and bridge-induced stability of an exemplary protein with $\theta$-curve motif and provide a catalogue of possible $\theta$-curves in proteins.
  \end{abstract}
  \end{@twocolumnfalse}
]

\begin{figure}[!t]
\begin{center}
\includegraphics[width=0.45\textwidth]{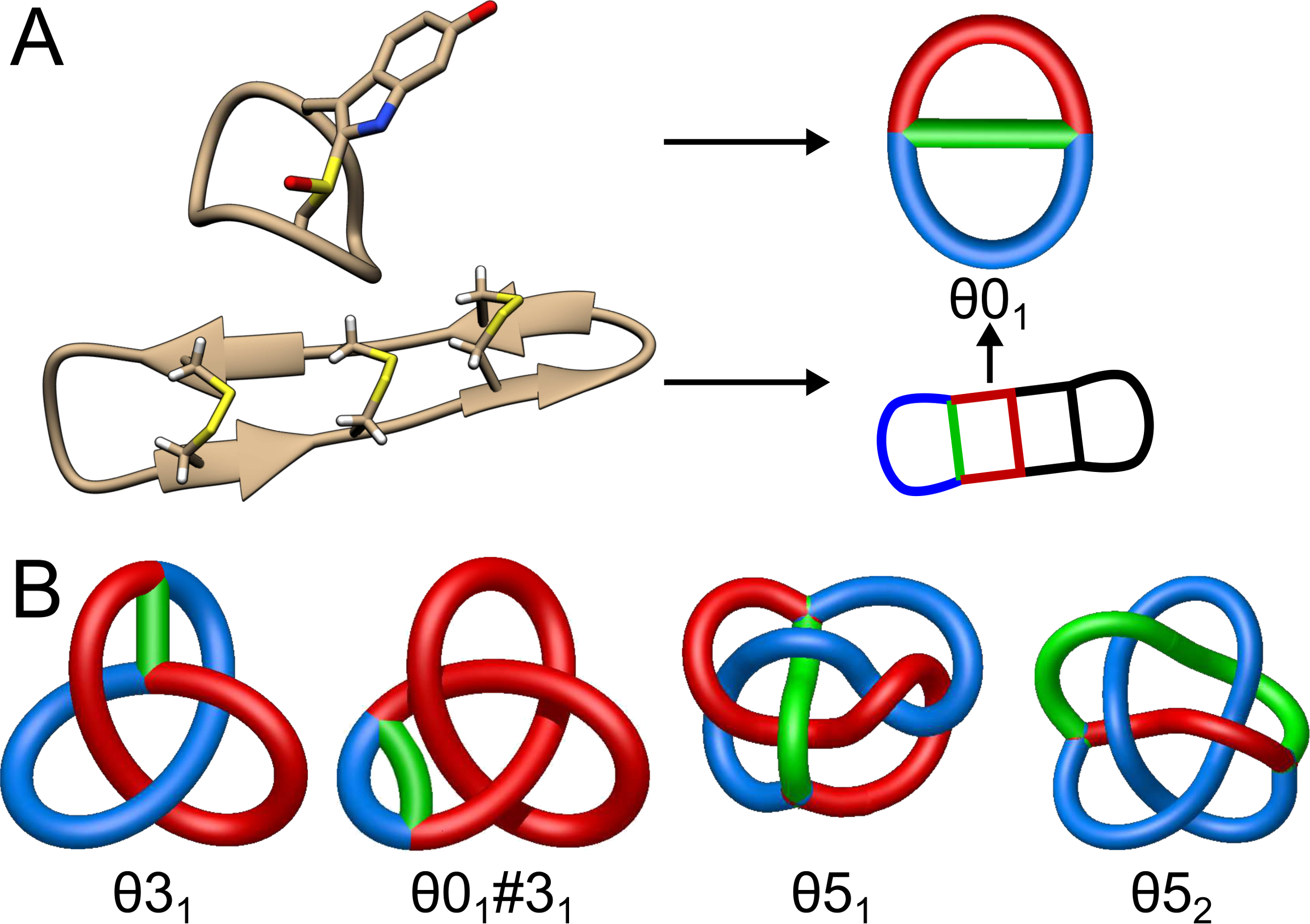}
\caption{\textbf{Simplest $\theta$-curves in proteins and classification of the $\theta$-curve motif}. (A) The $\alpha$-amantin (top, PDB code 6exvM) and the $\theta$-defensin (bottom, PDB code 2atg) -- cyclic oligopeptides with the bridges (shown explicitly) implying the existence of trivial $\theta$-curves ($\theta0_1$). In the case of $\theta$-defensin 3 bridges imply the existence of 10 different trivial motifs (one marked schematically). (B) The exemplary $\theta$-curves as classified in \cite{moriuchi2009enumeration}.\label{fig0}}
\end{center}
\end{figure}

Protein chains form numerous topologically complex structures, including knots  \cite{mansfield1994there,taylor2000deeply,nureki2002enzyme,nureki2004deep,bolinger2010stevedore,virnau2006intricate,dabrowski2017tie}, slipknots \cite{king2007identification,sulkowska2012conservation}, or links \cite{dabrowski2016lassoprot,caraglio2017physical,wikoff2000topologically}. Additionally, the covalent loops formed by the disulfide bridge may be pierced forming a lasso \cite{niemyska2016complex,dabrowski2016lassoprot,haglund2014pierced} or so-called cysteine knots \cite{craik1999plant,craik2001cystine}, or pierce one another, forming (deterministic) links \cite{boutz2007discovery,dabrowski2017topological,liang1995topological,liang1994knots}. Besides being appealing from the viewpoint of fundamental studies, such motifs were proved to be important for the protein's function or mechanical properties \cite{christian2016methyl,dabrowski2016search,san2017knots,sulkowska2008stabilizing,boutz2007discovery,zhao2017structural,haglund2017pierced,haglund2018uncovering, zhao2018stability}. However, the net of interactions resulting from the existence of intra-chain interactions creates much reacher world of topologically complex structures, some of which cannot be assigned to any class discussed. One of such motifs are $\theta$-curves (Fig.~\ref{fig0}).

The $\theta$-curves are embeddings of the Greek letter $\theta$ in the 3D space. Two embedding are equivalent if they can be transformed one into another by any continuous transformation, which does not involve the chain intersection. In proteins, the trivial $\theta$-curve motif (equivalent to the planar embedding of the letter $\theta$) is present within many proteins with internal bridges, e.g. in small, well-known proteins, such as $\alpha$-amantin or $\theta$-defensin (Fig.~\ref{fig0}A). It is, however, interesting to ask, if there are any non-trivial $\theta$-curves in proteins (Fig.~\ref{fig0}B)? Positive answer immediately results in a question about origin and function of such motif. On the other hand, if there is no non-trivial $\theta$-curve in proteins, one may wonder why and if there is any way to create them?

\begin{figure*}[t]
\begin{center}
\includegraphics[width=\textwidth]{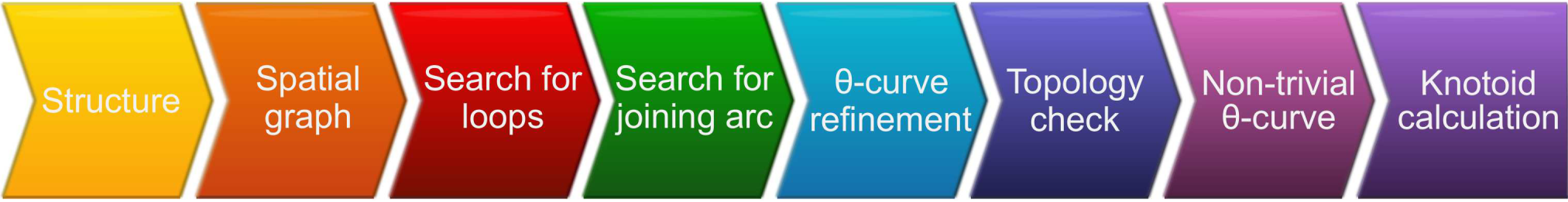}
\caption{\textbf{The identification method used in the work.} For the description of the individual steps, see the main text.\label{fig2}}
\end{center}
\end{figure*}

In this work, we challenge these questions. The paper is divided into four parts. The first one is devoted to the identification and classification of the non-trivial $\theta$-curve motif present in all known protein structures. In total, we identified 52 non-redundant structures, representing 7 topologically inequivalent (9 when counting chirality) non-trivial $\theta$-curves. To further specify the local $\theta$-curve geometry and distinguish between different $\theta$-curve spatial realization, we assign to each motif its  ``knotoid content'', i.e. the triple of knotoids (``open-chain knots'') constituting given $\theta$-curve.

In the second part, we utilize the classification to study the motif-function relation. We also analyze the motif conservation and organism of origin. In particular, we show the correlation with analogous features of main-chain knotted proteins and point to the cases, where the $\theta$-curve motif may not be accidental.

In the third part, we analyze the folding and the influence of the bridges on the stabilization of the protein, in comparison with analogous effects in proteins with main-chain knot and deterministic links. In particular, we show, that formation of the non-trivial structure in case of protein $\theta$-curve does not correlate with the folding barrier. In the last part we aim in creating a catalogue of all possible non-trivial $\theta$-curve motifs, which could be obtained either by introducing a bridge, or by synthesis of new, artificial proteins.

\section*{Results}
\subsection*{Identification and classification of proteins with the $\theta$-curve motif}
\subsubsection*{The algorithm}
The method used to identify the non-trivial $\theta$-curve topology is summarized in Fig. \ref{fig2}. First, the protein was represented as a spatial graph, spanned on the C$\alpha$ atoms representing each residue. The gaps (missing residues) were modelled as straight intervals. The graph edges represented either the protein backbone or bridge between residues. Three kinds of bridges were analyzed: 1) covalent (usually disulfide bridges), 2) ion-mediated (studied e.g. in \cite{liang1995topological,liang1994knots}), or 3) chain closure. In the case of chain closure, an additional vertex representing the point in infinity was added and connected to the terminal residues. The procedure of choosing a chain closure was performed 100 times and the dominating topology was chosen to represent the analyzed structure, consistently with the definition of the main-chain knot in proteins. Due to this method, the $\theta$-curves featuring bridge closure are called ``probabilistic'' through the rest of the article (as opposed to deterministic, not involving chain closure).

Next, we identified all cycles in a given graph (corresponding to the loops in proteins), and for each loop, we search for an arc (external to the loop) connecting two distinct residues in the loop. The $\theta$-curves obtained in such a way were validated, relaxed and simplified. In particular, we removed all structures with artificially long bonds or improbable gap filling. The topology of a $\theta$-curve was analyzed with three orthogonal methods implemented in the Topoly package \cite{dabrowski2019topoly}: calculation of the Yamada polynomial \cite{yamada1989invariant}, identification  of the Kauffman's boundary link analysis \cite{kauffman1993invariants}, and by calculation of the constituent knots (the knots formed by pairs of arcs). The analysis of the constituent knots was done as a validation of the results, as these knots do not distinguish well even the simplest $\theta$-curves. For example, both $\theta3_1$ and $\theta5_2$ (Fig.~\ref{fig0}B) have $3_1$, $0_1$, $0_1$ constituent knots, while $\theta5_1$ (Kinoshita curve) has only trivial constituent knots, just as like the trivial $\theta0_1$.

Finally, for each non-trivial $\theta$-curve, for its unsimplified, original structure, we calculated the dominating knotoid topology of each arc - the quantity which we call constituent knotoids (for the knotoid description see further parts of the manuscript). 

\subsubsection*{Types of $\theta$-curves in proteins}
\begin{figure*}[!t]
\begin{center}
\includegraphics[width=\textwidth]{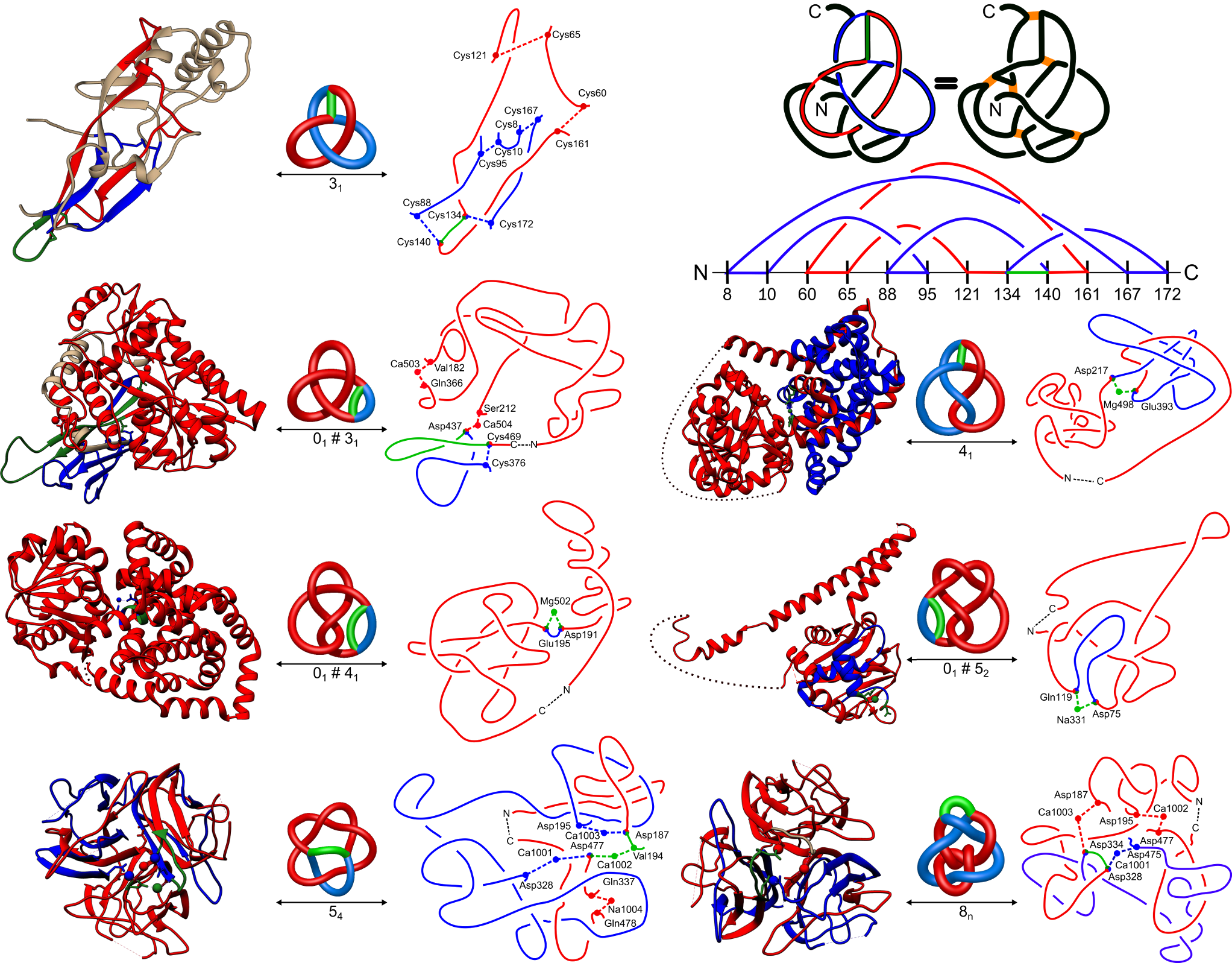}
\caption{\textbf{The identified $\theta$-curves.} For each motif, an exemplary structure is presented along with its simplification showing the actual topology. In the top-right corner a schematic depiction of the crab coagulogen chain forming the covalent $\theta3_1$ motif along with its bridge arrangement. The orange stripes denote the disulfide bridges. The dashed lines in the schemes denote the chain closure. The ``N'' and ``C'' letters denote the chain termini. The colours in the scheme match those in the structures. The PDB codes and structural details of presented structures are present in Tab \ref{tab-codes}\label{fig1}. For the many-chain $\theta$-curves see SI.}
\end{center}
\end{figure*}

As a result of our search, we identified non-trivial $\theta$-curves in 52 non-redundant protein chains (see Tab~\ref{tab-codes}) and 4 other cases, when the motif spanned two or three chains (connected by disulfide bridges). These structures represent 7 topologically different motifs (9 when counting chirality, see further parts of the text), presented schematically in Fig.~\ref{fig1}. In most cases these are probabilistic $\theta$-curves, but the motif present in the coagulogen from \textit{} with PDB code 1aoc is purely covalent (consisting of the pieces of backbone connected by disulfide bridges), and in 6 further one-chain cases and 1 two-chain case are ion-based $\theta$-curves (pieces of backbone connected either by disulfide bridge or interaction via ion).

\begin{table}[!t]
\begin{small}
\begin{tabular}{c|l}
\textbf{$\theta$-curve} & \textbf{PDB codes (nonredundant)} \\\hline
$\theta 3_1$ & \begin{tabular}{@{\extracolsep{\fill} }l}\textbf{1a8eA}, \textbf{1aocA}, \textbf{1aso}, 1aukA, \textbf{1b0l},\\
			1ei6A, 1ejjA, \textbf{1ihuA}, \textbf{1mdaM}, \textbf{1shqB},\\
			\textbf{2gsnB}, 2gsoA, 2hjnA, 2iucB, 2k0aA,\\
			\textbf{2oizD}, \textbf{2p4zA}, \textbf{2w5wB}, 2xrgA, \textbf{2z8fA},\\
			2zktA, \textbf{3b1bB}, \textbf{3c75L}, \textbf{3e2dA}, \textbf{3q3qA},\\
			3m8wA, \textbf{3qioA}, 3szyA, \textbf{3tg0B}, \textbf{3vysC},\\
			\textbf{4douA}, 4fdiA, \textbf{4fmwB}, \textbf{4hooB}, 4kayA,\\
			4lrdA, \textbf{5j81A}, \textbf{5ljzA}, \textbf{5osqA}, \textbf{3bpdCJ}\end{tabular} \\\hline
$\theta 0_1 \# 3_1$ & \begin{tabular}{@{\extracolsep{\fill} }l}1a42A, \textbf{1aocA}, \textbf{1aso}, 1aukA, \textbf{1b0l},\\
			\textbf{1dbiA}, \textbf{1ddzA}, 1ei6A, \textbf{1ihuA}, \textbf{1ru4A},\\
			\textbf{1shqB}, \textbf{2gsnB}, 2gsoA, 2hjnA, 2k0aA,\\
			\textbf{2p4zA}, \textbf{2w5wB}, 2xrgA, 2zktA, \textbf{3b1bB},\\
			\textbf{3e2dA}, 3m8wA, \textbf{3q3qA}, \textbf{3qioA}, 3szyA,\\
			\textbf{3tg0B}, \textbf{3vysC}, \textbf{4douA}, 4fdiA, 4h6vA,\\
			4lrdA, \textbf{4o6bB}, 5c74A, \textbf{5j81A}, \textbf{5ljzA},\\
			\textbf{5mqnA}, \textbf{5osqA}, \textbf{1fzcDEF}, \textbf{1lt9ABC}, \textbf{1re4DEF}\end{tabular} \\\hline
$\theta 4_1$ &  \begin{tabular}{@{\extracolsep{\fill} }l}3ulkA, 4wkkA\end{tabular}\\\hline
$\theta 0_1 \# 4_1$ &  \begin{tabular}{@{\extracolsep{\fill} }l}3ulkA, 4wkkA, 5e4rA, \textbf{1qmgA}, \textbf{3fr7A}\end{tabular} \\\hline
$\theta 0_1 \# 5_2$ &  \begin{tabular}{@{\extracolsep{\fill} }l}3ihrA\end{tabular}\\\hline
$\theta 5_4$ &  \begin{tabular}{@{\extracolsep{\fill} }l}\textbf{4douA}\end{tabular}\\\hline
$\theta 8_n$ &  \begin{tabular}{@{\extracolsep{\fill} }l}\textbf{4douA}\end{tabular}\\\hline
\end{tabular}
\caption{\textbf{PDB codes of non-redundant representative structures with non-trivial $\theta$-curves}. The capital letter denotes the chains forming the motif. Bold are the structures, which contain no main-chain knot nor slipknot.\label{tab-codes}}
\end{small}
\end{table}

In most cases (63\%, bold in Tab~\ref{tab-codes}) the $\theta$-curve motif arise within the proteins with unknotted backbone. Only the $\theta4_1$ and $\theta0_1\#5_2$ motifs are in each case the result of the main-chain knot. Apart from the motifs having a counterpart in the world of main-chain knotted proteins, human adiponectin hormone with PDB code 4dou forms two $\theta$-curves, which do not have any known analogue. One is $\theta5_4$ with $5_1$ constituent knot. The second one is 8-crossing $\theta$-curve with $8_5$ as one of its constituent knots. As the $\theta$-curves are classified up to 7 crossings this one remains unnamed, and therefore will be called $\theta8_n$. Both these $\theta$-curves feature the constituent knot with an unknotting number 2, never seen in proteins before. However, both cases seem to be rather the result of a dense net of interactions of four spatially close ions, possibly assigned automatically (the $\theta5_4$, features the highly dubious interaction Val-Ca). Apart from those two interesting cases, there are 2 prime $\theta$-curves ($\theta3_1$ and $\theta4_1$) and three composite ($\theta 0_1 \# 3_1$, $\theta 0_1 \# 4_1$ and $\theta 0_1 \# 5_2$). For the exemplary structures forming on-trivial $\theta$-curves see Tab. \ref{tab-arcs}.

\begin{table*}[!ht]
\begin{small}
\setlength{\tabcolsep}{2pt}
\begin{tabular}{c|c|c|l}
\textbf{$\theta$-curve} & \begin{tabular}{c}\textbf{PDB}\\\textbf{code}\end{tabular}& \textbf{Knotoid} & \multicolumn{1}{c}{\textbf{Arcs}}\\\hline

\multirow{3}{*}{$\theta 3_1$} & \multirow{3}{*}{1aocA} & 
	$k0_1$ & \begin{tabular}{l}C140 ... C161 $\leftrightarrow$ C60 ... C65$ \leftrightarrow$ C121 ... C134\end{tabular}\\
  &  & $k0_1$ & \begin{tabular}{l}C140 $\leftrightarrow$ C88 ... C95 $\leftrightarrow$ C10 ... C8 $\leftrightarrow$ C167 ... C172 $\leftrightarrow$ C134\end{tabular}\\
  &  & $k0_1$ & \begin{tabular}{l}C140 ... C134\end{tabular}\\\hline
  
 \multirow{3}{*}{$\theta 0_1 \# 3_1$} & \multirow{3}{*}{5osqA} & 
 	$k3_1$ & \begin{tabular}{l}D437 $\leftrightarrow$ Ca504 $\leftrightarrow$ S212 ... Q366 $\leftrightarrow$ Ca503 $\leftrightarrow$ V182 ... T3 $\leftrightarrow$ Cls $\leftrightarrow$ L474 ... C469\end{tabular}\\
 & & $k0_1$ & \begin{tabular}{l}D437 ... C376 $\leftrightarrow$ C469\end{tabular}\\
 & & $k0_1$ & \begin{tabular}{l}D437 ... C469\end{tabular}\\\hline
 
 \multirow{3}{*}{$\theta 4_1$} & \multirow{3}{*}{3ulkA} & 
 	$k0_1$ & \begin{tabular}{l}D217 ... M1 $\leftrightarrow$ Cls $\leftrightarrow$ V489 ... E393\end{tabular}\\
 & & $k0_1$ & \begin{tabular}{l}D217 ... E393\end{tabular}\\
 & & $k0_1$ & \begin{tabular}{l}D217 $\leftrightarrow$ Mg498 $\leftrightarrow$ E393\end{tabular}\\\hline

  \multirow{3}{*}{$\theta 0_1 \# 4_1$} & \multirow{3}{*}{5e4rA} & 
  	$k3_2$ & \begin{tabular}{l}D191 ... A2 $\leftrightarrow$ Cls $\leftrightarrow$ M476 ... E195\end{tabular}\\
 & & $k0_1$ & \begin{tabular}{l}D191 $\leftrightarrow$ Mg502 $\leftrightarrow$ E195\end{tabular}\\
 & & $k0_1$ & \begin{tabular}{l}D191 ... E195\end{tabular}\\\hline

\multirow{3}{*}{$\theta 0_1 \# 5_2$} & \multirow{3}{*}{3ihrA} & 
	$k5_2$ & \begin{tabular}{l}D75 ... E7 $\leftrightarrow$ Cls $\leftrightarrow$ L311 ... Q119\end{tabular}\\
 & & $k0_1$ & \begin{tabular}{l}D75 ... Q119\end{tabular}\\
 & & $k0_1$ & \begin{tabular}{l}D75 $\leftrightarrow$ Na331 $\leftrightarrow$ Q119\end{tabular}\\\hline
 
 \multirow{3}{*}{$\theta 5_4$} &\multirow{3}{*}{4douA} &
	  $k3_1$ & \begin{tabular}{l}D187 ... A108 $\leftrightarrow$ Cls $\leftrightarrow$ T525 ... Q478 $\leftrightarrow$ Na1004 $\leftrightarrow$ Q337 ... D477\end{tabular}\\
 & & $k0_1$ & \begin{tabular}{l}D187 $\leftrightarrow$ Ca1003 $\leftrightarrow$ D195 ... D328 $\leftrightarrow$ Ca1001 $\leftrightarrow$ D477\end{tabular}\\
 & & $k0_1$ & \begin{tabular}{l}D187 ... V194 $\leftrightarrow$ Ca1002 $\leftrightarrow$ D477\end{tabular}\\\hline
 
 \multirow{3}{*}{$\theta 8_n$} &\multirow{3}{*}{4douA} & 
	 $k0_1$ & \begin{tabular}{l}D334 $\leftrightarrow$ Ca1003 $\leftrightarrow$ D187 ... A108 $\leftrightarrow$ Cls $\leftrightarrow$ T525 ... D477 $\leftrightarrow$ Ca1002 $\leftrightarrow$ \\
	 	 $\quad\leftrightarrow$ D195 ... D328\end{tabular}\\
 & & $k0_1$ & \begin{tabular}{l}D334 ... D475 $\leftrightarrow$ Ca1001 $\leftrightarrow$ D328\end{tabular}\\
 & & $k0_1$ & \begin{tabular}{l}D334 ... D328\end{tabular}\\\hline
 \end{tabular}
\caption{\textbf{The structural details of proteins with exemplary $\theta$-curves shown in Fig.~\ref{fig1}}. The ``...'' denotes connection along the backbone, the $\leftrightarrow$ denotes the covalent bridge or interaction via ion. For the many-chain $\theta$-curves see SI.}
\label{tab-arcs}
\end{small}
\end{table*}

Although to our knowledge, the chirality of $\theta$-curves was not analyzed before, in the simplest cases, one can characterize their chirality by characterizing the chirality of its constituent knots. In particular, we observed both $\theta3_1$ and $\theta0_1\#3_1$ with both positive and negative trefoil constituent knot. Interestingly, however, the addition of a third arc transforms the achiral $4_1$ knot into chiral $\theta4_1$, as proven by nonsymetricity of its Yamada polynomial. Nevertheless, both proteins with $\theta4_1$ motif have the same $\theta$-curve chirality.

\subsubsection*{Multi-$\theta$-curve proteins and knotoid analysis}
\begin{figure}[!h]
\begin{center}
\includegraphics[width=0.5\textwidth]{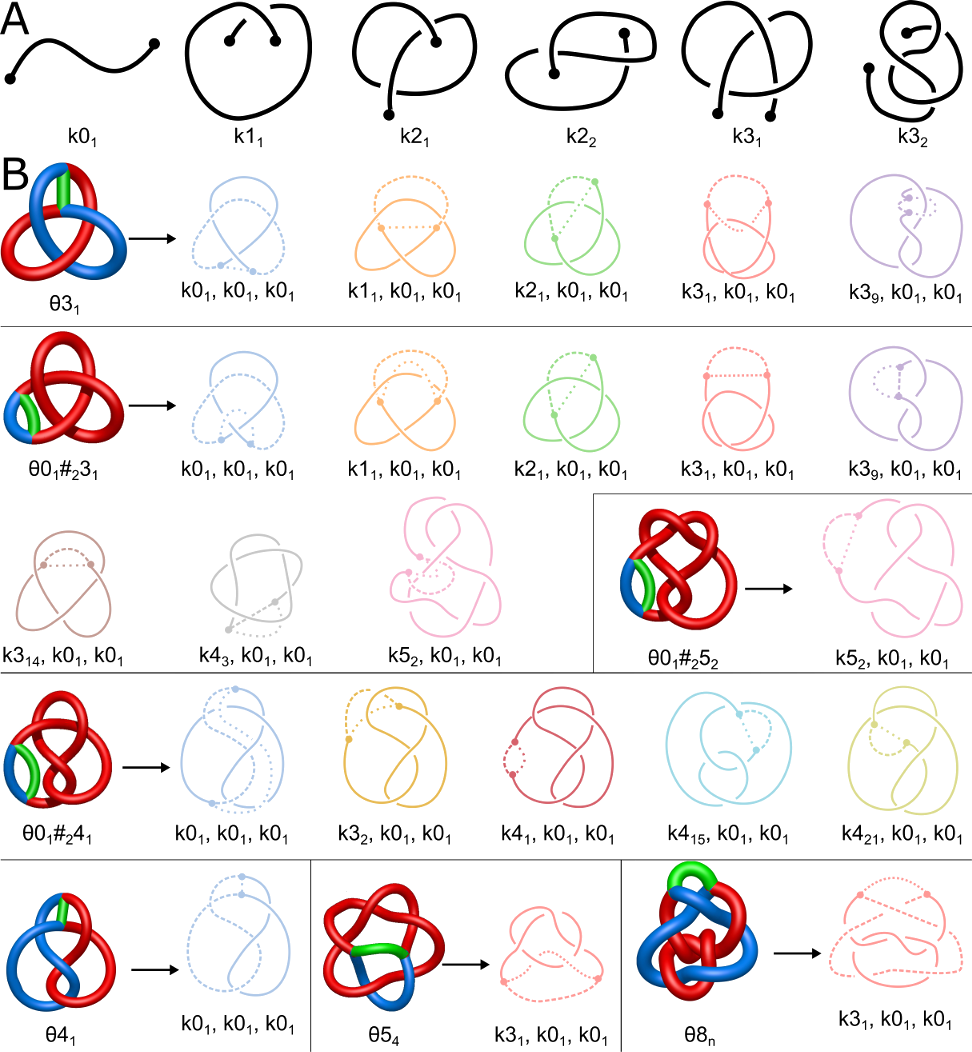}
\caption{\textbf{Constituent knotoids of identified $\theta$-curves.} The colours denote the topology of the most complicated knotoid (marked with solid line). The knotoid notation comes from \cite{dimos}.\label{fig3}}
\end{center}
\end{figure}

The number of non-trivial $\theta$-curves in one chain depends strongly on the number of ions and covalent bonds, reaching up to 70 different, non-trivial $\theta$-curves in viral protein with PDB code 5j81. In principle, proteins with a different number of $\theta$-curves and motifs represented may differ also in biological function and physical properties. Moreover, the $\theta$-curve of the same topology may differ in the  realization of the motif (see Fig. \ref{fig3}), which can also influence the protein's features. In order to characterize the $\theta$-curve spatial realization to some extent, we turned towards the knotoid approach \cite{turaev2012knotoids}. 

The knotoids are planar arcs with information about over- and undercrossings. They knotoid graphs are regarded as equivalent, if they can be transformed one into another with a continuous transformation, keeping the surrounding of the termini intact. For a 3D curve, we perform a large set of different projections and prescribe the dominating knotoid type. Such approach was already used to study the local knot conformation of proteins \cite{goundaroulis2017studies,goundaroulis2017topological}. In this case, we describe each non-trivial $\theta$-curve by its constituent knotoids -- a triple of knotoid types corresponding to arcs forming $\theta$-curve.

The set of knotoid contents determined for identified $\theta$-curves is presented in Fig.~\ref{fig3}. The greatest variety of $\theta$-curve realizations is visible for $\theta0_1\#3_1$ motif. In some cases it seems, that slightly different connection of the trivalent points would result in structures in very interesting topology (e.g. in case of $\theta 0_1 \# 3_1$ with $k5_2$ knotoid).

\subsection*{Function, origin and conservation of the $\theta$-curve motif}
The extensive classification of the $\theta$-curves allowed us to study the topology-function correlation and origin of the motif. In Tab.~\ref{tab-function} we co-located the motifs of representative structures with their function and kingdom of an organism of origin. To extract only the features possibly stemming from the deterministic knot or the $\theta$-chain motif, Tab.~\ref{tab-function} contains only the main-chain unknotted structures.

\begin{table}[!hb]
\begin{small}
\setlength{\tabcolsep}{2pt}
\begin{tabular}{c|l|l}
\textbf{$\theta$-curve} & \textbf{Functions} & \textbf{Kingdom}\\\hline
$+\theta 3_1$ &
		 \begin{tabular}{@{\extracolsep{\fill} }l}Oxidored. (2), adhesion (2),\\
			viral (2), transport (2),\\
			hydrolase, transferase,\\
			coagulation, metal\\
			binding, hormone\end{tabular}  
		& \begin{tabular}{@{\extracolsep{\fill} }l}Bacterie (6),\\Animals (5),\\Viruses (3)\end{tabular}\\\hline
		
$-\theta 3_1$ & 
		\begin{tabular}{@{\extracolsep{\fill} }l}Hydrolase (5), isomerase,\\transport, coagulation,\\binding \end{tabular}
		&\begin{tabular}{@{\extracolsep{\fill} }l} Bacteria (6),\\Archaea,\\Animals \end{tabular}\\\hline
		
$\theta 0_1 \# +3_1$ 
		& \begin{tabular}{@{\extracolsep{\fill} }l}Lyase (4), viral (2),\\
			adhesion (2), hydrolase (2),\\
			hormone, metal binding\end{tabular} 
		&\begin{tabular}{@{\extracolsep{\fill} }l}  Animals (6),\\Viruses (3),\\Bacterie (3),\\Rhodophytas,\\Archaea\end{tabular} \\\hline

$\theta 0_1 \# -3_1$ 
		& \begin{tabular}{@{\extracolsep{\fill} }l}Hydrolase (7), isomerase,\\transport, coagulation\end{tabular} 
		& \begin{tabular}{@{\extracolsep{\fill} }l}Bacterie (7),\\Archaea,\\Animals\end{tabular} \\\hline

$\theta 0_1 \# 4_1$ 
		& \begin{tabular}{@{\extracolsep{\fill} }l}Oxidored. (2) \end{tabular}
		&\begin{tabular}{@{\extracolsep{\fill} }l} Plants (2)\end{tabular} \\\hline

$\theta 5_4$ 
		& \begin{tabular}{@{\extracolsep{\fill} }l}Hormone\end{tabular} 
		& \begin{tabular}{@{\extracolsep{\fill} }l}Animals\end{tabular} \\\hline

$\theta 8_n$ 
		&\begin{tabular}{@{\extracolsep{\fill} }l} Hormone\end{tabular} 
		& \begin{tabular}{@{\extracolsep{\fill} }l}Animals\end{tabular} \\\hline

\end{tabular}
\caption{\textbf{Function and origin of proteins with $\theta$-curves}.\label{tab-function}}
\end{small}
\end{table}

The analysis of the Tab.~\ref{tab-function} shows, that around 70\% of structures with $\theta$-curves are enzymes, although the enzymes constitute only around 30\% of all known structures. This resembles the case of main-chain knotted proteins, for which it was hypothesized, that the non-trivial topology helps in the formation of places favourable for enzymatic active sites \cite{dabrowski2016search}. Possibly the same mechanism is present in case of enzymes with non-trivial $\theta$-curves. 

Closer analysis shows, that the function of the proteins is correlated with the constituent knot, rather than with the whole $\theta$-curve motif. The exception are the lyases with $\theta0_1\#3_1$ motif. Interestingly, these falls in the category of carbohydrate anhydrases, which other family members feature shallowly knotted main-chain. This again shows that the knotted loop may perform a similar function as the main-chain knot.

The analysis of the organisms of origin shows interesting correlation of higher organisms (animals) with positive chirality of the constituent knot, while there are only singular animal representants of structures with negative trefoil. This again turns our attention towards the animal carbonic anhydrases with $\theta0_1\#3_1$ motif. On the other hand, there is no visible preference of formation of any $\theta$-curve motif in bacteria or viruses, which seems to indicate, that these motifs could appear accidentally.

The question of the functionality of the motif can be also approached by analyzing the conservation of the motif. Namely, high conservation of the motif among the homologs indicates its importance for the protein. In such a way one may show the importance of the main-chain knot or deterministic links in proteins cite{sulkowska2012conservation,dabrowski2017topological}. The conservation of $\theta$-curve motif relies crucially on the conservation of the bridges. In general, proteins with at least 50\% of sequential similarity have all the pivotal, motif-forming residues conserved, however, below this threshold, some crucial disulfide bonds or ion-binding residues are lost. This does not provide any valuable information about the importance of the motif. Conversely, there are proteins, for which the $\theta$-curve motif is unique among all its homologs. For example, the oxidoreductase from \textit{Cucurbita pepo} with PDB code 1aso compared to its homologs has additional disulfide bond, required for the $\theta$-curve formation. In this case, the $\theta$-curve motif is therefore rather accidental and do not provide any particular function for the protein.

\subsection*{Folding and bridge induced stability}
The folding of main-chain knotted proteins invariably requires threading of the main chain, which is believed to be a rate-determining step \cite{sulkowska2009dodging,a2013folding,li2014energy}. To overcome the barrier, some external factors or alternative pathways were suggested \cite{chwastyk2015cotranslational,dabrowski2018protein,lim2015mechanistic}. We analyzed folding of the only covalent $\theta$-curve protein -- Japanese horseshoe crab coagulogen (PDB code 1aoc) \cite{bergner1996crystal} -- tracking also the formation of the knotted cycle. We performed a series of constant-temperature folding/unfolding simulations in coarse-grained C$\alpha$ model to calculate the free energy landscape of the protein. To check the influence conditions, we mimicked the oxidative and reductive conditions by the strenght of Cys-Cys non-bonding interaction. It turns out, that in both cases the free energy landscape is very similar and increasing the disulfide bridges strength results only in slight lowering of the free energy barrier.

\begin{figure}[!htb]
\begin{center}
\includegraphics[width=0.48\textwidth]{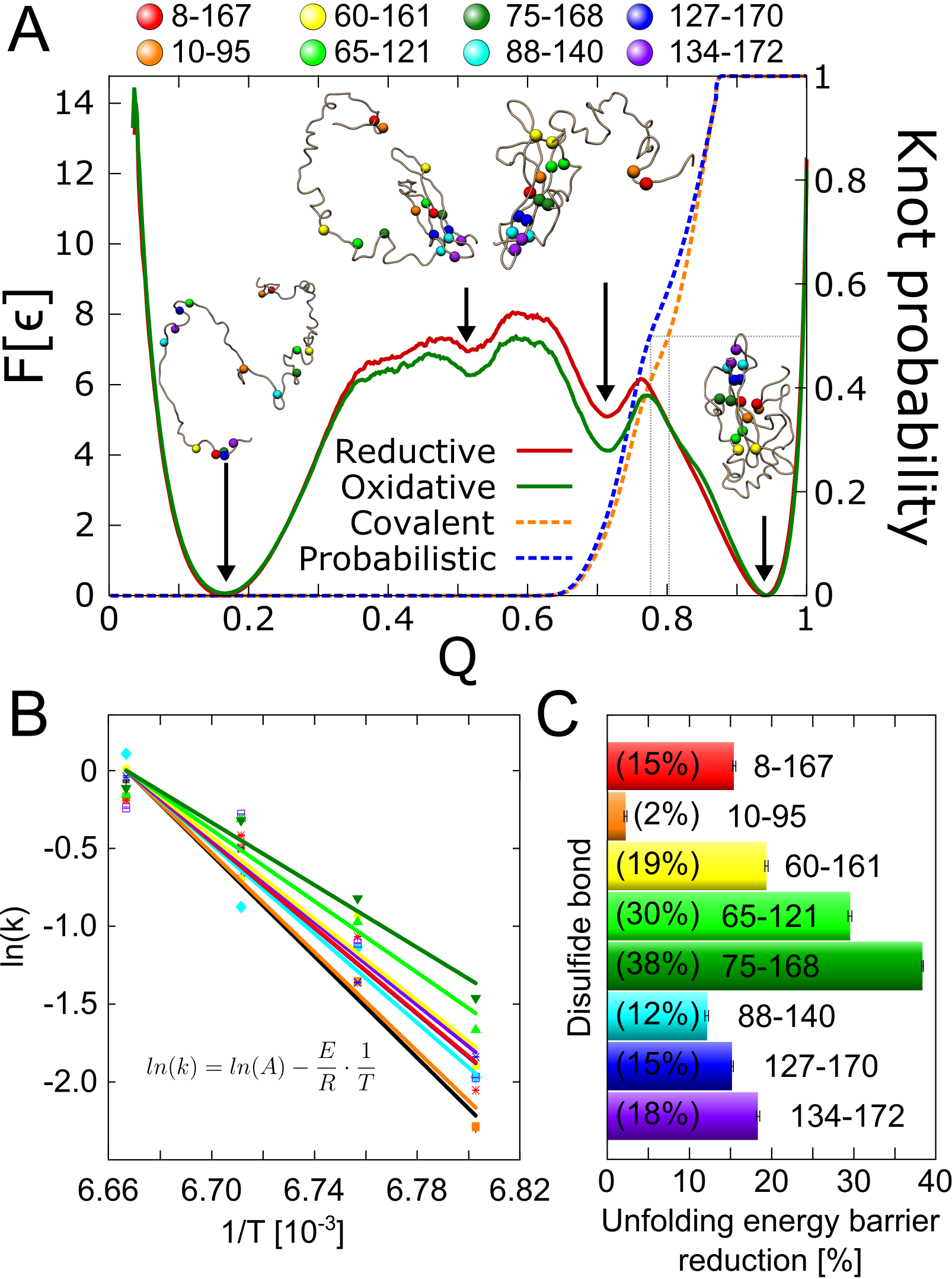}
\caption{\textbf{Folding and bridge induced stability analysis.} (A) The free energy landscape of the crab coagulogen (PDB code 1aoc) with the probability of knotted loop formation overlayed. Apart from two minima corresponding to the unfolded and folded state, the folding pathway features the existence of two local minimas corresponding to intermediate products $I_1$ and $I_2$. On top, the colour code of the beads' indices. The dashed lines indicate the probability of knotted loop formation equal to 50\%. (B) The logarithm of the unfolding rate as the function of the reverse of temperature for models with one bridge missing. (C) The influence of the removal of one bridge on the free energy barrier on unfolding, as compared to the original structure.\label{fig5}}
\end{center}
\end{figure}

In both cases, the folding follows the path with formation of two short-living intermediate products (called $I_1$ and $I_2$), differing in the set of disulfide bonds formed (Fig. \ref{fig5}A). In the $I_1$ structure three C-terminal disulfide bonds (indices 88-140, 127-170, and 134-172) building the cysteine knot motif are formed. The passage $I_1\rightarrow I_2$ requires travesting over the maximum of the free energy barrier, when three additional bridges are created (indices 60-161, 65-121, and 75-168). Creation of the N-terminal bonds (indices 8-167 and 10-95) related to passing over small local maximum of free energy barrier completes the folding. The creation of these bonds is necessary to form a knotted loop (both deterministic or probabilistic). Therefore, the knot is formed in late stages of folding, for fraction of native contacts $Q\sim0.8$. This is very different compared to the case of main-chain knots, where the formation of a knot corresponds to the maximum of free energy barrier and occurs much earlier in folding (for $Q\leq0.6$). This indicates, that the formation of the knotted loop is in this case not a crucial point on the folding pathway and is rather a result of the result of the tertiary structure.

Another feature of proteins with main-chain knot and deterministic links is the stability induced by the complex topology. To quantify the stabilization effect of the knot forming bridges we performed the constant temperature unfolding simulations for the models with one bridge removed. Our reasoning was, that if any knotted loop is more important in the stabilization of the protein structure, removing its disulfide bridges should result in destabilization of the structure, visible in the speed of thermal unfolding. For each model, we calculated the unfolding rate constant and presented it as a function of the inverse of temperatur (Fig. \ref{fig5}B). Fitting this to Arhenius equation allowed us to calculate the energy barrier on unfolding (normalized by gas constant), as the slope of curve. This allowed us to calculate the change in the barrier height, as a fraction of the barrier height for the original protein (Fig. \ref{fig5}C, numerical data in SI). 

It turns out, that the most stabilization is induced by bridges with indices 75-168, 134-172, 60-161, and 8-167. Three of the bridges are part of the covalent knotted loop, but only two of them are parts of all three knotted loops present in this protein (one deterministic and two probabilistic, involving chain closure). Two of the most stabilizing bridges are fixing the termini of the protein, and two more are protecting the core of the protein from falling apart. On the other hand, the deterministic loop includes also the bridge 10-95, which add the least stabilization for the protein. Therefore it seems, that the protein stabilization results from holding together various pieces of the chain, which accidentally results in a knotted loop formation. It does not, however, indicate that the knotted loop itself is a key stabilizing factor for the structure.

\subsection*{The catalogue of possible $\theta$-curves in proteins}
In previous sections, we presented the set of all non-trivial $\theta$-curves present in proteins. However, the motif strongly depends on the presence of bridges (covalent and ion-based). This in principle creates the possibility of designing new knotted loops and $\theta$-curve by point mutations introducing additional bridges. Therefore, we wanted to check, if any other knotted loops or $\theta$-curve motifs can be obtained in this way.

\begin{figure}[!ht]
\begin{center}
\includegraphics[width=0.45\textwidth]{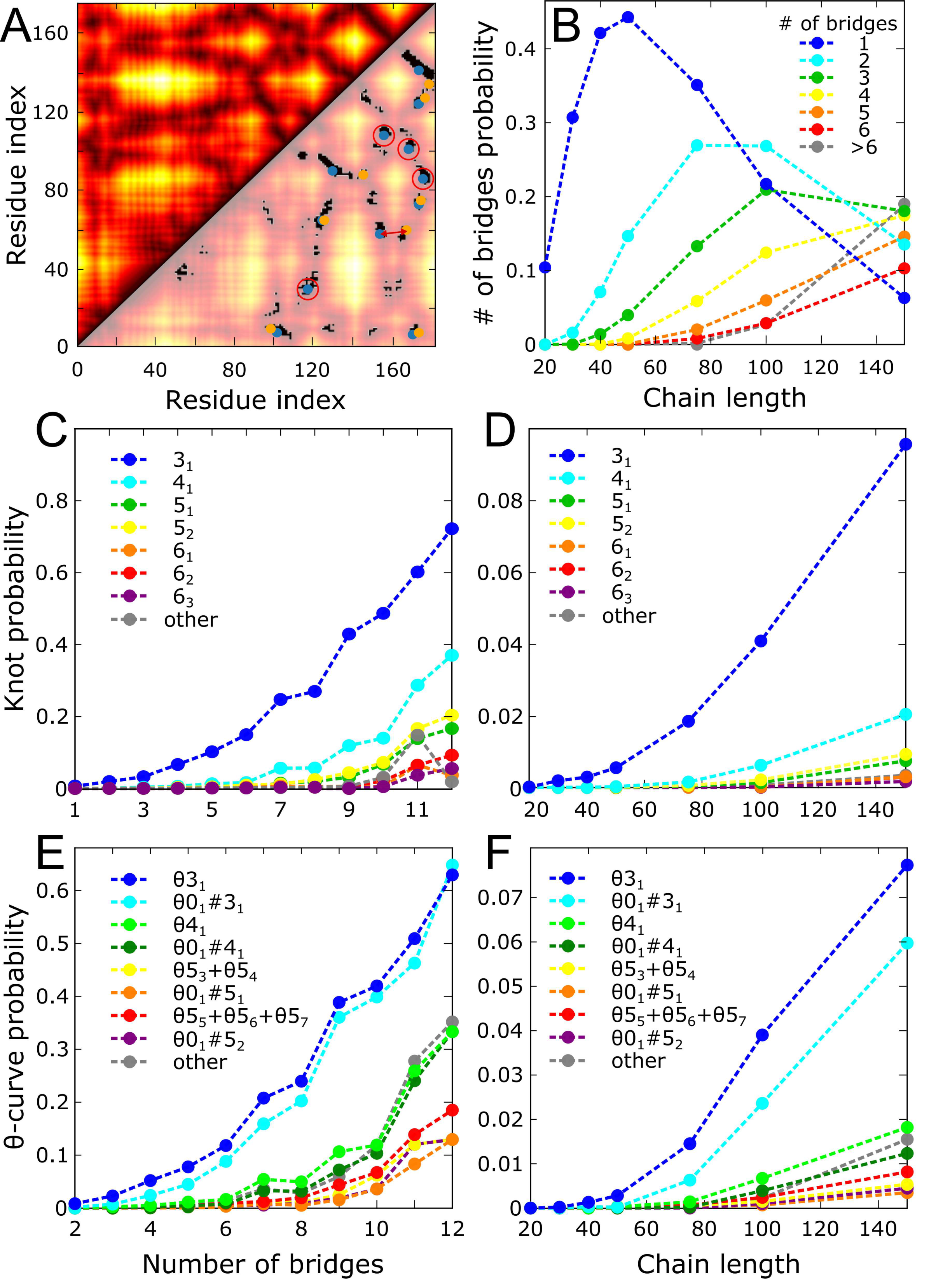}
\caption{\textbf{The analysis of possible protein knotted loops and $\theta$-curves.} (A) The method of obtaining the proposed bridges. From the contact map the residue pairs within cutoff are chosen (black spots under diagonal), for which the representative pair is chosen (blue dot). This correspond to the original bonds (orange dots), with for additional bridges (encircled red). (B) The number of proposed bridges as a function of polymer length. (Middle row) The statistical probability of obtaining knotted loop as a function of (C) number of bridges number and (D) chain length. (Bottom row) The statistical probability of obtaining non-trivial $\theta$-curve as a function of (E) number of bridges and (F) chain length.\label{fig7}}
\end{center}
\end{figure}

The introduction of the bridge is usually done by substituting the residues, which are spatially close in the original structure. We singled out the groups of such contacts, and for each group, we chose one representant (for the detailed description of the method see SI). The representative contacts serve as our candidates for the bridge (blue dots in Fig. \ref{fig7}A). In total, we usually find not less than the number of bridges present in the original structure, and each original bridge (orange dot in Fig. \ref{fig7}A) has its corresponding proposed bridge. This technique allows us to predict the set of proposed bridges for any structure, which is not smaller than the set of original bridges. Therefore, to check if by point mutation any new motifs can be obtained, we scanned the whole non-redundant set of proteins, for each of them adding the proposed bridges.

As a result, we found no new $\theta$-curve motif, which means, that introduction of a new bridge into any known structure would not lead to a new deterministic $\theta$-curve motif. However, one could obtain a new probabilistic $\theta$-curve motif by introducing a bridge into $5_2$ or $6_1$ main-chain knotted proteins. For example, when considering the $6_1$ knotted hydrolase with PDB code 3bjx, introducing a bridge connecting residues 132 and 281 (or some in nearest vicinity) results in probabilistic $\theta6_6$, while bridging residues 47-170 results in probabilistic $\theta6_5$. Similarly, introducing a bridge 3-87 to the $5_2$  knotted hydrolase ligase results in probabilistic $\theta5_6$ structure.

The result that no new deterministic $\theta$-curves may be found in proteins by introducing a bridge is rather surprising. It suggests, that the set of possible $\theta$-curves in proteins may be somehow restricted by the protein internal geometry. In fact, the Ramachandran angles induce some non-uniform distribution on the planar and torsion angles between consecutive C$\alpha$ atoms. Therefore we asked a more general question, what $\theta$-curve motif can the protein chain attend? Or in other words, what motifs could be synthesized by expressing some human-invented, artificial sequence? This question is similar in nature to the question of what knotted compounds can be obtained using the template synthesis \cite{polles2015self,marenda2018discovering}. To answer this question, we created a very simple equilateral polymeric model with the correct distribution of planar and dihedral angles, mimicking the protein chain. The distribution was obtained by calculating the angles for all proteins in the non-redundant protein set (see SI). Next, we chose 7 different chain length, and for each length we generated 10.000 structures (70.000 chains in total). For each chain, we generated the set of proposed bridges and we searched for the non-trivial $\theta$-curves.

The number of bridges identified increased with the chain length (Fig. \ref{fig7}B), reaching, for the longest polymers singular cases with 15 bridges. The number of bridges is critical for the presence of non-trivial topology. In particular, only for structures with 7 or more bridges the statistical probability of obtaining a deterministic knot is larger than 0.2. In accordance, the only protein with purely covalent knot (crab coagulogen with PDB code 1aoc) has 8 bridges. For large number of bridges (>10) the probability of obtaining the detrministic knot is significant, especially, if the brides are clustered alltogether. Therefore, even for the long chains, if the chain is not compact enough, allowing for creation of large number of bridges, the probability of obtaining knotted loop is low.

As expected, the dominating non-trivial deterministic knotted loop is the $3_1$ knot, followed by $4_1$ and $5_2$. However, the $5_1$ knot has a statistical probability comparable with $5_2$. This is an unexpected result, especially as the $5_1$ knot has unknotting number 2, therefore it needs at least two threadings to be formed. However, in case of knotted loops it may be much easier to form, compared to main-chain knot. One way is to start from a structure with $3_1$-knotted main chain and complete the remaining threading by disulfide bridge. Similar reasoning was already proposed \cite{flapan2015topological} and in fact, the $5_1$ knotted cycle, based on the interaction via ions is present in protein with PDB code 4dou.

The composition of possible $\theta$-curves is a direct consequence of the set of knotted loops identified. The most probable are $\theta3_1$ and $\theta0_1\#3_1$, followed by $\theta4_1$ and $\theta0_1\#4_1$. The next are $5_2$-based $\theta$-curves ($\theta5_5$, $\theta5_6$, $\theta5_7$, and $\theta0_1\#5_2$) and $5_1$-based $\theta$-curves ($\theta5_3$, $\theta5_4$ and $\theta0_1\#5_1$). However, there are also structures containing both $3_1$ and $4_1$ knotted loops (like $\theta6_4$), and other more complex cases. Although individually each structure is rather improbable, in total the ``other'' group has statistical probability comparable to the $\theta4_1$. This shows, that new $\theta$-curve types could be present rather in the proteins with a highly clustered net of disulfide interactions.

\section*{Discussion}
The concept of knotted loops including disulfide bridges and interactions via ion a has long history, but only moderate results were published \cite{liang1994knots,crippen1974topology,klapper1980knotting,benham1993disulfide}. Only recently, with over 100.000 protein structures known, the first classes of topologically non-trivial structures based on disulfide bridges -- links and lassos -- were defined. In this work, we added another small brick - the non-trivial $\theta$-curves.

As a result of the scan of all known structures, we identified 52 non-redundant one chain and 4 many-chain proteins with $\theta$-curve motif, representing 7 (9 including chirality) different topologies.

The classification of $\theta$-curves allowed us to scrupulously compare the topology and the function of the proteins with $\theta$-curves. The result, that most of the $\theta$-curve containing proteins are enzymes indicates, that the non-trivial topology may induce the chain rigidity, which in turn may create places favourable for enzymatic active sites, similarly as in the case of main-chain knotted proteins \cite{dabrowski2016search}. An example can be the carbohydrate anhydrases, whose some homologs have a shallow $3_1$ main-chain knot.

Looking at the organism of origin, it seems, that the $\theta$-curve structures are more possible in the lower organisms, such as bacteria or viruses and no preference towards any chirality, or topology may be identified. On the other hand, there is a strong preference for animal proteins, to attain the positive chirality in the case of $3_1$ based $\theta$-curves. Therefore one may speculate, that the $\theta$-curves present in lower organisms may occur on random, and the animal structures are the ones which could be functional. However, the determination of the functional advantage of complex topology seems a hard task, being elusive also for main-chain knot.

To analyze the possible function of the $\theta$-curve motif, we analyzed the folding and stability of the only purely covalent $\theta$-curve. The analysis showed no clear signs of the special importance of the knot-related bridges for the stability. Also, conversely to the main-chain knot in proteins, the formation of the non-trivial structure based on disulfide bonds do not correlate with the maximum of the free energy profile. This shows, that possibly the formation of the knotted loops and the $\theta$-curves may be much easier to achieve, than knotted proteins. Taking into account the possile function of knotted loops in the carbohydrate anhydrases it seems, that the knotted loops may be an easier-to-achieve way to obtain at least some features of the main-chain knot. Such a perspective creates a question, which $\theta$-curves may be formed in proteins.

Our analysis of this issue showed, that the deterministic $\theta$-curves are possible rather in the structures with a dense net of inter-residue covalent interactions. In priciple, it would be interesting, to analyze the physical properties of a protein-based $\theta$-curves and relate them to the topology, similarly, as the hydrodynamical properties of circular DNA chains were related to their knot topology \cite{weber2013sedimentation,vologodskii1998sedimentation}. Some results on the hydrodynamical properties of $\theta$-curve polymers were already calculated \cite{deguchi2017statistical}. Moreover, one could also confine the protein chain in order to increase the knot probability. Such a technique was used to explain the problem of main-chain knotted protein folding \cite{zhao2018exclusive,soler2016steric} and to increase the probability of knot formation in polymers \cite{d2017linking}. 

This work can be continued in many directions. From the mathematical side, the $\theta$-curves are only the simplest structures with many-valent vertices. One could classify the whole protein graph created by all disulfide bridges and interaction via ions at once. In principle, this is possible by assigning e.g. the Yamada polynomial for each graph. However, the most challenging would be to finding the biological relevance in such data. Secondly, a deeper understanding of the relation between knotted loop topology and the function could lead to designing proteins with desired properties. Finally, the physical properties of proteins and polymers, in general, need to be established. Still, much work can be done.

\section*{Methods}
\textbf{Protein dataset} All the protein structures (128378) deposited in PDB until March 2018 were used. The protein gaps were filled with a straight interval.
\\\textbf{Topology determination} The topology of $\theta$-curves was determined using the tools from Topoly package~\cite{dabrowski2019topoly}, utilizing also the HOMFLY-PT polynomial for knot determination ~\cite{freyd1985new,przytycki2016invariants,ewing1991load}. Knotoids were analyzed with Knoto-ID software~\cite{dorier2018knoto}. The knotoid notation comes from \cite{dimos}. The structures were presented as spatial graphs and the loops were searched using Python implementation of DFS algorithm.
\\\textbf{Folding simulations} The simulations were performed using Gromacs 4.5.4 with Gaussian potential for non-bonding interactions and parameters proposed by SMOG server.
\\\textbf{Structure presentation} The structures were visualized using UCSF Chimera~\cite{pettersen2004ucsf}. 

\section*{Acknowledgements}
This work was supported by National Science Centre \#2012/07/E/NZ1/01900 (to JIS) and \#2016/21/N/NZ1/02848 (to PDT), EMBO Installation Grant \#2057 (to JIS), Leverhulme Trust \#RP2013-K-017 (to DG and AS) and by the Swiss National Science Foundation \#31003A-138267 (to DG and AS).

\bibliography{biblio}
\bibliographystyle{naturemag}

\end{document}